\documentclass[12pt]{iopart}
\usepackage[latin1]{inputenc}
\usepackage[T1]{fontenc}
\usepackage{iopams}
\usepackage{graphicx}
\usepackage{cite}
\usepackage{siunitx}
\usepackage{tikz}
\usepackage{subfigure}

\usepackage{color}

\begin{document}
\sisetup{
	retain-explicit-plus = true
}
\rapid{Bridge connection of quantum Hall elementary devices}
\author{Luca Callegaro\textsuperscript{1} and Massimo Ortolano\textsuperscript{1,2} 
		\\[\medskipamount]
        	{\small \textsuperscript{1}INRIM - Istituto Nazionale di Ricerca Metrologica} \\
        	{\small Strada delle Cacce, 91 - 10135 Torino, Italy} \\		
		{\small \textsuperscript{2}Dipartimento di Elettronica e Telecomunicazioni, Politecnico di Torino} \\
        	{\small Corso Duca degli Abruzzi 24, 10129 Torino, Italy} \\
}

\begin{abstract}
Multiple-series and multiple-parallel connections of quantum Hall elementary devices allow the realization of multiple or fractional values of the quantized Hall resistance, rejecting the effect of contact and wiring resistances. We introduce here the \emph{multiple-bridge} connection, which maintains the properties of multiple-series and parallel connections and allows more freedom in the choice of the topology of networks composed of quantum Hall elements, and the design of more efficient quantum Hall array resistance standards (and other devices). As an example, a $5$-element network is analyzed in detail.
\end{abstract}

\maketitle
\section{Introduction}
Electrical networks composed of interconnected multiterminal quantum Hall effect (QHE) devices are of great interest for electrical metrology. Quantum Hall array resistance standards (QHARS)~\cite{Piquemal:1999,Poirier:2002,Bounouh:2003,Poirier:2004,Hein:2004,Oe:2008,Oe:2011,Woszczyna:2012,Oe:2013} are integrated devices which allow the representation of the ohm in the International System of Units with values (e.g., decadic) suitable for the calibration of artifact resistance standards of practical interest; other networks realize intrinsically-referenced voltage dividers~\cite{Domae:2012} or allow universality tests~\cite{Schopfer:2007} with unprecedented sensitivity levels. 

All such networks are presently based on the repeated application of the so-called \emph{multiple-series} or \emph{multiple-parallel} connections~\cite{Delahaye:1993}. The denomination of these connections is taken from the usual series or parallel connection of two-terminal resistors. Their purpose is to realize a compound network of two QHE elements, having a four-terminal resistance value which is close to twice (for multiple-series) or half (for multiple-parallel) the resistance $R_\textup{H}$ of each element, with a strong rejection of the unavoidable resistances of the contacts and of the electrical wiring. 

\begin{figure}[htb]
	\centering
	\includegraphics[width=3cm]{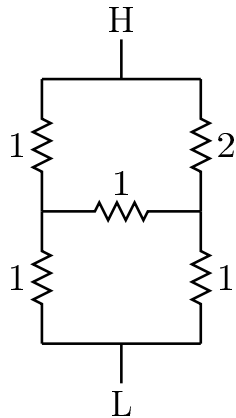}%
 	\caption{Two-terminal resistors arranged in a bridge network.  With the normalized values given in the figure, the resistance between terminals H and L is $R_\textup{HL} = 13/11$. \label{fig:basic_bridge}}%
\end{figure}

A generic electrical network of two-terminal resistors not only includes series or parallel connections; a major building block is, in fact, the \emph{bridge} connection. Figure~\ref{fig:basic_bridge} shows a network composed of two-terminal resistors, connected in a bridge configuration. 

In this Communication, we investigate the realization of a \emph{multiple-bridge} connection of QHE elements. We show that the multiple-bridge connection has, in common with the multiple-series and the multiple-parallel ones, the property of strong rejection of contact and wiring resistances. The multiple-bridge connection can be employed as a building block of QHE networks; in particular, it allows the design of QHARS having values that, if realized only with double-series or double-parallel connections, would require a larger number of QHE elements; hence, it can improve the design of QHARS of practical interest.

\section{Preliminaries and notation}
\begin{figure}
\centering
	\subfigure[]{\includegraphics{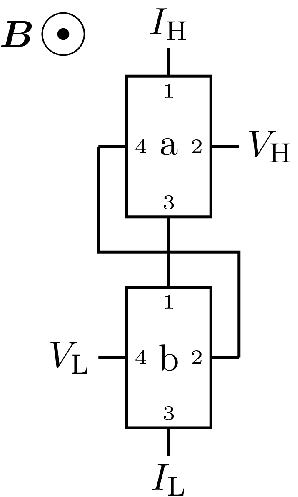} \label{fig:double_series}}
	\subfigure[]{\includegraphics{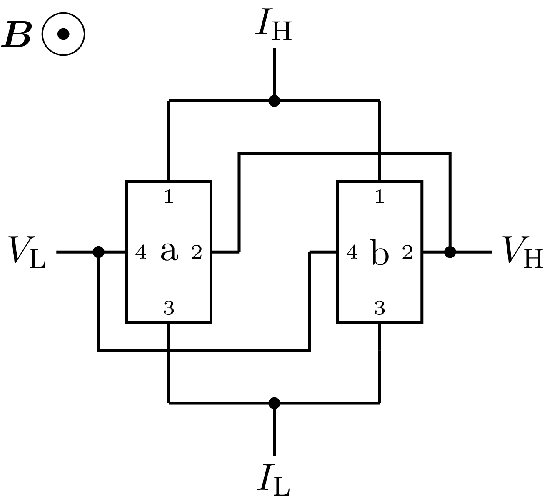} \label{fig:double_parallel}}
\caption{\subref{fig:double_series} Double series and \subref{fig:double_parallel} double parallel connection of two quantum Hall elements. Each element is provided with two current contacts ($1$,$3$) and two voltage contacts ($2$,$4$). The resulting networks have two current terminals ($I_\textup{H}, I_\textup{L}$) and two voltage terminals ($V_\textup{H}, V_\textup{L}$) each.}
\end{figure}

In what follows, QHE elements are labelled with lower-case letters. The magnetic
flux density $\bi{B}$ is assumed to be pointing out of the page. All QHE elements are supposed to be in the same fully quantized state, with $R_\textup{H} = R_\textup{K}/i$, where $R_\textup{K}$ is the von Klitzing constant and $i$ is the plateau index. We employ the notation given below, partly taken from~\cite{Delahaye:1993,Ortolano:2012}:

\begin{itemize}
\item $r = R_\textup{H} / 2$ is the resistance associated with the Ricketts-Kemeny model~\cite{Ricketts:1988} of a QHE element;
\item $\epsilon_{k\textup{x}}\, r$ is the resistance of contact $k$ (including the wiring resistance) of element x;
\item $m$ is the connection order. $m=2$ corresponds to a double-series (figure~\ref{fig:double_series}), double-parallel (figure~\ref{fig:double_parallel}) or double-bridge connection; $m=3$ corresponds to the triple-series, triple parallel or triple-bridge connection;
\item $R_\textup{S}^{(m)}$ is the four-terminal resistance of the $m$-series connection of two QHE elements;
\item $R_\textup{P}^{(m)}$ is the four-terminal resistance of the $m$-parallel connection of two QHE elements.
\end{itemize}

It was shown in~\cite{Delahaye:1993,Cage:1998,Ortolano:2012} that
\begin{equation}
R_\textup{S}^{(m)} = 2 R_\textup{H} \left( 1 + O(\epsilon^m) \right), \qquad  R_\textup{P}^{(m)}  = \frac{1}{2} R_\textup{H} \left( 1 + O(\epsilon^m) \right),
\end{equation}
where $O({\boldsymbol{\cdot}})$ is the \emph{big O notation}, the order of magnitude of the dependence on $\epsilon$.


\section{Double-bridge connection}
\begin{figure}[htb]
	\centering
	\includegraphics[width=7cm]{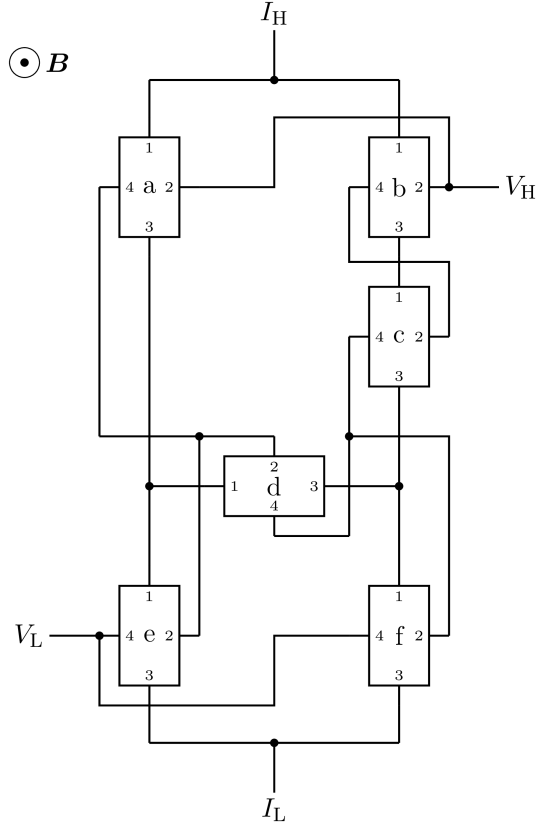}
	\caption{A network of five QHE elements, with a topology corresponding to that of figure~\ref{fig:basic_bridge} and including a double-bridge connection. The elements are provided with two current contacts ($1$,$3$) and two voltage contacts ($2$,$4$). The network has two current terminals ($I_\textup{H}, I_\textup{L}$) and two voltage terminals ($V_\textup{H}, V_\textup{L}$).}
\label{fig:doublebridge}
\end{figure}

Figure~\ref{fig:doublebridge} shows the connection of five QHE elements in order to realize a four-terminal network with a topology corresponding to that given in figure~\ref{fig:basic_bridge}. 
The four-terminal resistance $R_\textup{B}^{(2)}$ of the network has been computed with the matrix analysis method described in~\cite{Ortolano:2012}:
\begin{eqnarray}\label{eq:double_bridge}
R_\textup{B}^{(2)} &= \frac{13}{11} R_\textup{H} \Bigg(1+\frac{49 \epsilon_{1\textup{a}} \epsilon_{2\textup{a}}}{1144}-\frac{7 \epsilon_{1\textup{a}} \epsilon_{2\textup{b}}}{286}-\frac{7 \epsilon_{1\textup{b}} \epsilon_{2\textup{a}}}{286}+\frac{2 \epsilon_{1\textup{b}} \epsilon_{2\textup{b}}}{143} \nonumber \\ 
&\qquad+\frac{2 \epsilon_{1\textup{c}} \epsilon_{2\textup{c}}}{143}+\frac{2 \epsilon_{1\textup{c}} \epsilon_{4\textup{b}}}{143}+\frac{\epsilon_{1\textup{d}} \epsilon_{2\textup{d}}}{858}-\frac{\epsilon_{1\textup{d}} \epsilon_{2\textup{e}}}{286} \nonumber \\
&\qquad+\frac{7 \epsilon_{1\textup{d}} \epsilon_{4\textup{a}}}{1716}-\frac{\epsilon_{1\textup{e}} \epsilon_{2\textup{d}}}{286}+\frac{6 \epsilon_{1\textup{e}} \epsilon_{2\textup{e}}}{143}+\frac{7 \epsilon_{1\textup{e}} \epsilon_{4\textup{a}}}{286} \nonumber \\
&\qquad+\frac{25 \epsilon_{1\textup{f}} \epsilon_{2\textup{f}}}{858}+\frac{5 \epsilon_{1\textup{f}} \epsilon_{4\textup{c}}}{429}+\frac{5 \epsilon_{1\textup{f}} \epsilon_{4\textup{d}}}{1716}+\frac{2 \epsilon_{2\textup{c}} \epsilon_{3\textup{b}}}{143} \nonumber \\
&\qquad+\frac{7 \epsilon_{2\textup{d}} \epsilon_{3\textup{a}}}{1716}+\frac{7 \epsilon_{2\textup{e}} \epsilon_{3\textup{a}}}{286}+\frac{5 \epsilon_{2\textup{f}} \epsilon_{3\textup{c}}}{429}+\frac{5 \epsilon_{2\textup{f}} \epsilon_{3\textup{d}}}{1716} \nonumber \\
&\qquad+\frac{49 \epsilon_{3\textup{a}} \epsilon_{4\textup{a}}}{858}+\frac{2 \epsilon_{3\textup{b}} \epsilon_{4\textup{b}}}{143}+\frac{8 \epsilon_{3\textup{c}} \epsilon_{4\textup{c}}}{429}-\frac{\epsilon_{3\textup{c}} \epsilon_{4\textup{d}}}{429} \nonumber \\
&\qquad-\frac{\epsilon_{3\textup{d}} \epsilon_{4\textup{c}}}{429}+\frac{\epsilon_{3\textup{d}} \epsilon_{4\textup{d}}}{858}+\frac{9 \epsilon_{3\textup{e}} \epsilon_{4\textup{e}}}{286}-\frac{15 \epsilon_{3\textup{e}} \epsilon_{4\textup{f}}}{572} \nonumber \\ 
&\qquad-\frac{15 \epsilon_{3\textup{f}} \epsilon_{4\textup{e}}}{572}+\frac{25 \epsilon_{3\textup{f}} \epsilon_{4\textup{f}}}{1144}\Bigg)\,, \\
&= \frac{13}{11} R_\textup{H} \left( 1 + O(\epsilon^2) \right). 
\end{eqnarray}
Equation~\eref{eq:double_bridge} shows that the double-bridge connection allows the rejection of the parasitic resistances to the order $O(\epsilon^2)$, characteristic of the double-series and double-parallel connections.
\section{Triple-bridge connection}
\begin{figure}[htb]
	\centering
	\includegraphics[width=7cm]{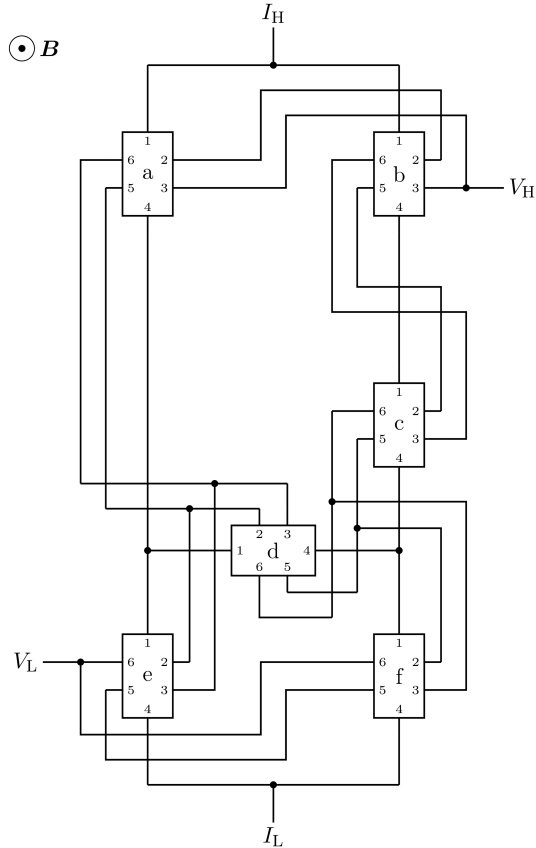}
	\caption{The same network of figure~\ref{fig:doublebridge}, here in triple-bridge connection. Each element is now provided with two current contacts ($1$,$4$) and four voltage contacts ($2$,$3$,$5$,$6$). \label{fig:triplebridge}}%
\end{figure}

Multiple-bridge connections having order $m>2$ can be realized. Figure~\ref{fig:triplebridge} shows a triple-bridge connection of six-terminal QHE elements, having the same topology of figures~\ref{fig:basic_bridge} and~\ref{fig:doublebridge}. The resulting four-terminal resistance has a convoluted expression\footnote{To those interested, the authors can provide the Mathematica\textsuperscript{\textregistered} notebook of the full calculation.}, which however can be summarized as $R_\textup{B}^{(3)} = \frac{13}{11} R_\textup{H} \left( 1 + O(\epsilon^3)\right)$, as expected from triple series or parallel connections.
\section{Conclusions}
We have shown that the multiple-bridge connection of quantum Hall elements maintains the rejection properties of multiple-series and multiple-parallel connections. The inclusion of the multiple-bridge connection as a building block allows the realization of new QHE networks, which may result more efficient than the ones including only multiple-series or multiple-parallel connections. This is demonstrated by the example given in figures~\ref{fig:basic_bridge}, \ref{fig:doublebridge} and \ref{fig:triplebridge}, where the normalized value $13/11$ is achieved with $5$ QHE elements; the same value requires at least $8$ elements to be generated in networks without bridges. Whether a similar efficiency improvement can be achieved in devices of practical interest, such as decadic-valued QHARS, will be the subject of future investigations.

\section*{References}
\bibliographystyle{iopart-num}
\bibliography{../QHE}
\end{document}